\documentclass[sigconf]{acmart}
\usepackage{booktabs}
\usepackage{multirow}
\usepackage{graphicx}
\usepackage{subcaption}
\usepackage{amsmath}
\usepackage{algorithm}
\usepackage{algpseudocode}
\usepackage{dblfloatfix}
\usepackage{xspace}
\usepackage{enumitem}
\usepackage{placeins}

\algrenewcommand\algorithmicrequire{\textbf{Input:}}
\algrenewcommand\algorithmicensure{\textbf{Output:}}

\newcommand{\sys}{RedMem-Store\xspace}

\setcopyright{none}
\acmDOI{}
\acmISBN{}
\acmConference[Conference 'XX]{the International Conference on Conference}{2027}{}
\acmYear{2027}
\copyrightyear{2027}

\begin{document}
\settopmatter{printacmref=true}

\title{\sys: Characterizing and Optimizing Storage Engine for LLM User Memory at Production Scale}

\author{Bing Tian}
\authornote{Work done during internships at Xiaohongshu, Inc.}
\affiliation{%
  \institution{Huazhong University of Science and Technology}
  \city{Wuhan}
  \country{China}
}
\email{tbing@hust.edu.cn}
\author{Xiao Chen}
\affiliation{%
  \institution{Xiaohongshu Inc.}
  \city{Shanghai}
  \country{China}
}\email{chenxiao4@xiaohongshu.com}

\author{Haikun Liu}
\correspondingauthor 
\affiliation{%
  \institution{Huazhong University of Science and Technology}
  \city{Wuhan}
  \country{China}
}\email{hkliu@hust.edu.cn}

\author{Rui Ma}
\affiliation{%
  \institution{Xiaohongshu Inc.}
  \city{Shanghai}
  \country{China}
}\email{marui4@xiaohongshu.com}

\author{Yiping Sun}
\affiliation{%
  \institution{Xiaohongshu Inc.}
  \city{Shanghai}
  \country{China}
}\email{sunyiping@xiaohongshu.com}

\author{Jinwei Zhang}
\affiliation{%
  \institution{Xiaohongshu Inc.}
  \city{Shanghai}
  \country{China}
}\email{zhangjinwei1@xiaohongshu.com}

\author{Xiangrui Yang}
\affiliation{%
  \institution{The University of Hong Kong}
  \city{Hong Kong}
  \country{China}
}\email{yxr620@gmail.com}

\author{Zhuohui Duan}
\affiliation{%
  \institution{Huazhong University of Science and Technology}
  \city{Wuhan}
  \country{China}
}\email{zhduan@hust.edu.cn}

\author{Lipeng Liang}
\affiliation{%
  \institution{Xiaohongshu Inc.}
  \city{Shanghai}
  \country{China}
}\email{gaosaroma@gmail.com}

\author{Fengjun Zhang}
\affiliation{%
  \institution{Xiaohongshu Inc.}
  \city{Shanghai}
  \country{China}
}\email{yuancin@163.com}

\author{Yue Chen}
\affiliation{%
  \institution{Xiaohongshu Inc.}
  \city{Shanghai}
  \country{China}
}\email{chenyue14@xiaohongshu.com}

\author{Xiaocheng Zhong}
\affiliation{%
  \institution{Xiaohongshu Inc.}
  \city{Shanghai}
  \country{China}
}\email{zhongxiaocheng@xiaohongshu.com}

\author{Zhiyong Wang}
\affiliation{%
  \institution{Xiaohongshu Inc.}
  \city{Beijing}
  \country{China}
}\email{sunzhenghuai@xiaohongshu.com}

\author{Xiaofei Liao}
\affiliation{%
  \institution{Huazhong University of Science and Technology}
  \city{Wuhan}
  \country{China}
}\email{xfliao@hust.edu.cn}


\begin{abstract}
Memory systems enable LLM applications to personalize question answering, recommendation, and semantic search using information retained from previous interactions. At RedNote (a.k.a.\ Xiaohongshu), a global social networking platform serving 400 million users, the workload is heavily skewed: the vast majority of users hold only a few memories and often remain idle. Existing memory frameworks store memories in general-purpose vector databases, posing a fundamental scalability dilemma: isolating each user incurs prohibitively high storage overhead, whereas pooling all users into a shared index significantly increases query I/Os to the global corpus rather than the target user's data.

We present \sys, a storage engine for production-scale user memory that
resolves this dilemma through a \emph{virtual collection} abstraction,
providing logical per-user isolation over physically shared storage. This
design eliminates per-user fixed storage overhead while keeping  memory usage
proportional to active users and confining each request's I/O to individual user data. Specifically, \sys incorporates three novel designs: 1) an \emph{access-aware virtual collection layout} to locate active users in memory and read only required disk blocks, 2) \emph{user-batched in-place updates} to coalesce small writes and amortize update I/Os, and 3) \emph{in-collection hybrid retrieval} to filter vectors and search keywords over shared ordinals, deferring the read of texts in users' memory to mitigate unnecessary I/Os. Across production-scale workloads, \sys preserves P99 query latency at about 1\,ms regardless of the number of users.  Compared with Qdrant and Milvus, \sys reduces P50 query latency by up to 25.8$\times$ and 14.8$\times$, and increases write throughput by up to 11.7$\times$ and 6.8$\times$, respectively. \sys has been deployed  and  served 400 million users stably for several months, significantly  reducing the hardware cost of RedNote's production memory service by over 90\%.

\end{abstract}

\renewcommand{\shortauthors}{Tian et al.}
\maketitle

\section{Introduction}
\label{sec:intro}

\begin{figure}[t]
  \centering
  \includegraphics[width=\columnwidth]{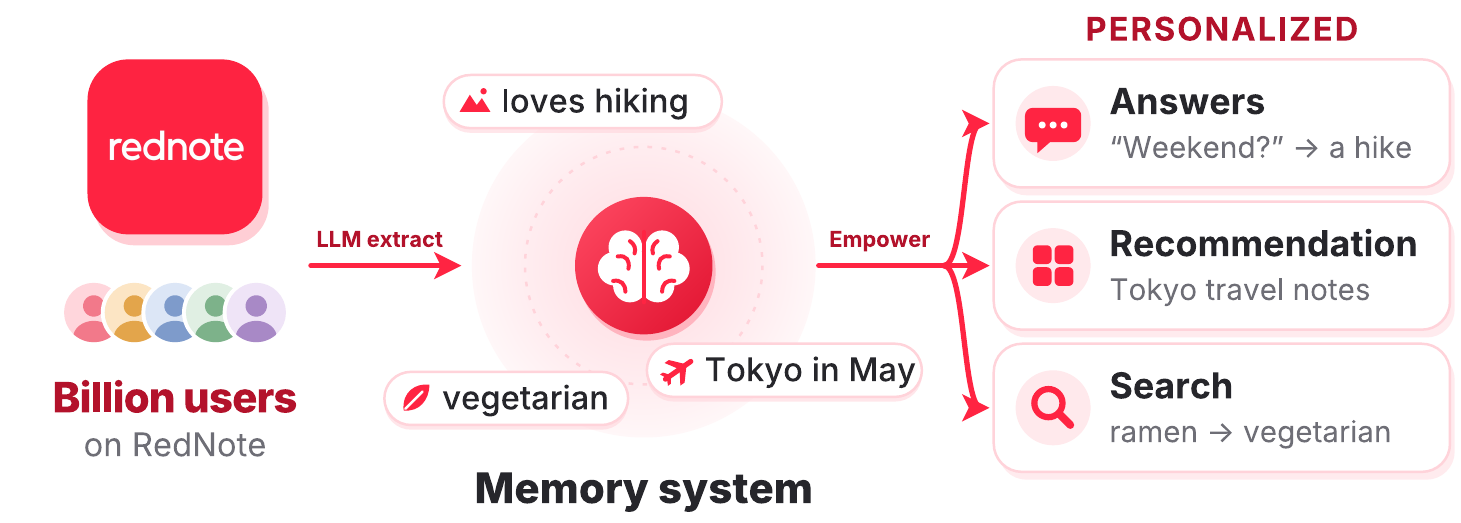}
  \caption{Memory systems personalize question answering,
  recommendation, and search for RedNote's users.}
  \label{fig:overview}
\end{figure}

Large language model (LLM) applications increasingly rely on memory systems to retain user-specific context across sessions and personalize responses~\cite{packer2023memgpt,zhang2024survey,chhikara2025mem0,li2025memos}. At RedNote, a global social networking platform serving 400 million users, the user memory system serves as a foundational building block for LLM-driven question answering, recommendation, and semantic search. As shown in Figure~\ref{fig:overview}, each personalized request first retrieves the user's most relevant memories and then injects them into the model prompt. Because this retrieval sits on the critical serving path, the underlying storage engine has become the primary determinant of both tail latency and infrastructure cost.


Recent memory frameworks such as Mem0~\cite{chhikara2025mem0} and MemOS~\cite{li2025memos} focus on \emph{what} to remember, i.e., extracting, organizing, and updating memories from interactions. For \emph{where} to store them, however, they rely on general-purpose vector databases such as Qdrant~\cite{qdrant} and Milvus~\cite{wang2021milvus}. In practice, deploying these databases for multi-user memory forces a choice between two organizations: an \emph{isolated} layout that allocates one dedicated collection per user, or a \emph{shared} layout that pools all user memories into a single collection and scopes queries via a user-ID filter.


Our production-scale experience demonstrates that neither organization is cost-efficient at RedNote's scale (\S\ref{sec:motivation}). Under the isolated layout, the storage overhead scales with the \emph{total user count} rather than the actual memory volume. For users holding at most ten memories, the physical footprint inflates to 8,327$\times$ the logical data size, and collection creation fails at roughly 10,000 users on a single node due to file-descriptor exhaustion (Figure~\ref{fig:isolation}). Conversely, the shared layout eliminates this per-user fixed cost, but bounds the query cost to the \emph{global corpus}. When we expand a shared collection from 100,000 to 400 million users while keeping the queried users and their memories fixed, the average read I/O per query significantly increases 25.4$\times$, from 566\,KiB to 14.0\,MiB. Under replayed production traffic, the P99 latency spikes around 500 \textit{queries per second} (QPS) and reaches 102\,ms at 800\,QPS.


These bottlenecks stem from a structural mismatch: general-purpose vector databases are engineered for corpus-oriented retrieval-augmented generation (RAG), whereas production user memory should be designed for ultra-scale multi-tenancy scenarios~\cite{aulbach2008multitenant} with hundreds of millions of small, mostly-idle tenants (\S\ref{sec:workload}). Unlike RAG which ingests a few large corpora in bulk and searches across them, the user memory system often partitions data into hundreds of millions of tiny per-user datasets, absorbs a continuous stream of small user-scoped writes, and serves queries strictly within a single user's footprint. Because vector databases manage storage, segments, and indexes at the collection level, mapping each user to a collection replicates corpus-scale machinery per user, while mapping each user to a filter forces every query to suffer from corpus-wide metadata and segment fragmentation. Even recent multi-tenancy optimizations, such as co-locating tenant records inside segments or hashing tenants into shared partitions~\cite{qdrantmultitenancy,milvusmultitenancy}, still retain global segments and partitions as the fundamental units of storage and search.

Drawing on our production characterization, we argue that a storage engine for user memory must decouple its costs from both the total user population and the global corpus size: the disk footprint should scale with stored memories, the DRAM usage should be proportional to active users, and per-request I/O should be strictly confined to the queried user's data.

We present \sys, a storage engine designed around these principles through a \emph{virtual collection} abstraction: each user views a logically private collection with its own identifier space, keyword statistics, and access paths, while all collections are physically packed into shared pages and tracked via a compact resident directory and on-demand block maps. We propose three mechanisms to make this architecture practical at scale. 1) An \emph{access-aware virtual collection layout} (\S\ref{sec:storage}) locates individual users via a single in-memory lookup, caches block maps only for active users, separates quantized codes, postings, metadata, and text by access paths, and packs small collections into shared pages. 2) To handle continuous small writes, \emph{user-batched in-place updates} (\S\ref{sec:update}) coalesce writes per user, apply each user's batch atomically in place, and delegate durability to the upstream Kafka log rather than maintaining a local write-ahead log. 3) On the read path, \emph{in-collection hybrid retrieval} (\S\ref{sec:retrieval}) evaluates metadata filters, vector similarity, and keyword relevance strictly within the target user's collection, merges candidate results via shared ordinals, and materializes full text payloads only for the final top-$k$ records.

\sys has been successfully deployed as RedNote's production memory service for several months. We evaluate \sys on production memories with 400 thousand to 400 million users and compare to Qdrant and Milvus, each configured with its recommended multi-tenancy settings (\S\ref{sec:eval}). As the number of users grows by 1,000$\times$, \sys keeps its P50 query latency within 0.49--0.73\,ms and its P99 latency within 0.91--1.12\,ms. Compared with Qdrant and Milvus, it reduces P50 query latency by 5.1--25.8$\times$ and 8.0--14.8$\times$, reduces P99 latency by 7.0--23.8$\times$ and 11--590$\times$, and increases write throughput by 3.2--11.7$\times$ and 2.2--6.8$\times$, respectively. Moreover, \sys  significantly reduces the hardware cost of RedNote’s production memory service by over 90\%, saving total 12,000 CPU cores and 92\,TB DRAM.

In summary, this paper makes the following contributions:
\begin{itemize}[leftmargin=1.5em]
  \item \textbf{Workload characterization and bottleneck analysis.} We characterize the workload of RedNote's 400-million user memories  and empirically identify the root causes behind why both isolated and shared vector-databases layouts fail to scale (\S\ref{sec:motivation}).
  \item \textbf{Virtual-collection storage engine.} We design and implement \sys, which achieves per-user cost proportionality through an access-aware virtual collection  layout, user-batched in-place updates, and in-collection hybrid retrieval (\S\ref{sec:design}).
  \item \textbf{Evaluation and production lessons.} We demonstrate the superiority of \sys over state-of-the-art vector database---Qdrant and Milvus using production workloads and standard memory benchmarks, and distill operational lessons learned from deploying it at RedNote (\S\ref{sec:eval}--\S\ref{sec:lessons}).
\end{itemize}
\section{Background}
\label{sec:background}


\subsection{Memory Systems for LLM Applications}
\label{sec:bg-memory}

A memory system retains what an LLM application learns about each user and supplies relevant context at serving time~\cite{packer2023memgpt,zhang2024survey}. At the storage layer, a \emph{memory} is a short record owned by a single user, consisting of raw text, an embedding vector, and metadata such as type (e.g., preference, fact, or event) and timestamps. The storage engine supports two user-scoped operations. \emph{Writes} insert, update, or delete memories for a given user. \emph{Retrievals}, denoted $\mathrm{Retrieve}(u, q, F, k)$, take a user $u$, query $q$, and metadata filter $F$, returning the top-$k$ memories of $u$ ranked by vector similarity to $q$ and, optionally, keyword relevance such as BM25~\cite{robertson2009bm25}. The application then packs these records into the LLM prompt alongside the user request.

State-of-the-art memory frameworks such as Mem0~\cite{chhikara2025mem0} and MemOS~\cite{li2025memos} diverge mainly in memory management: Mem0 prompts an LLM to extract and reconcile candidate memories against existing ones, while MemOS exposes memory as schedulable, versioned \emph{MemCubes} modeled on OS resources. Beneath this management layer, both frameworks share the same storage architecture---offloading persistence and retrieval to an off-the-shelf vector database, optionally paired with a graph store for entity relations. RedNote's memory service follows this same design.

\subsection{Vector Databases}
\label{sec:bg-vdb}

Because memory retrieval hinges on approximate nearest-neighbor (ANN) search over embeddings, vector databases serve as the default storage backend. Consider Qdrant~\cite{qdrant}, a backend supported by both Mem0 and MemOS. Qdrant organizes data into \emph{collections}, each partitioned into \emph{segments} that maintain independent vector stores, payload (metadata) stores, and indexes such as HNSW graphs~\cite{malkov2018hnsw}. Incoming writes land in small mutable segments, which a background optimizer periodically merges and indexes. At scale, a single collection accumulates hundreds of segments. Every search executes across all segments in the collection and merges the per-segment candidates into the final top-$k$. Milvus~\cite{wang2021milvus} adopts a similar segment-oriented pipeline within a disaggregated architecture: writes enter growing segments that are sealed and indexed once full, and queries fan out across all segments before merging results.

To restrict retrieval to a single user, vector databases rely on filtered search over payload fields. Given a payload index on the user ID, Qdrant's query planner estimates the filter's cardinality within each segment: if few points match, it scores them exactly; otherwise, it traverses the segment's HNSW graph while applying the filter on the fly. Critically, both planning and execution operate at segment granularity.



\begin{figure}[t]
  \centering
  \includegraphics[width=\columnwidth]{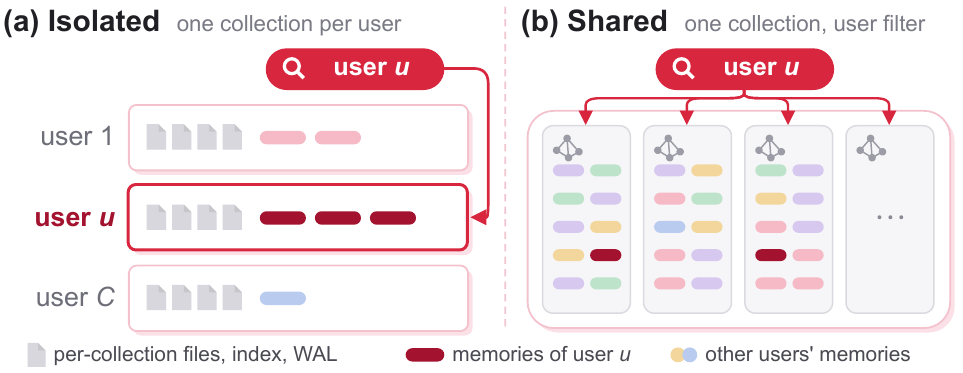}
  \caption{Two organizations for storing user memories in vector databases.}
  \label{fig:organizations}
\end{figure}

\subsection{Storing Memories at Production Scale}
\label{sec:bg-org}

Serving many users from one database is a classic multi-tenancy problem, spanning from dedicated instances or schemas per tenant to shared tables~\cite{aulbach2008multitenant}. Existing memory frameworks sit at the two extremes of this spectrum, as shown in Figure~\ref{fig:organizations}. The \emph{isolated} organization assigns a separate collection to each user. Queries route directly to that user's collection without a user-ID filter, but every user carries the full machinery of a collection---segments, indexes, a write-ahead log, and metadata configuration. Conversely, the \emph{shared} organization pools all users into a single collection and isolates queries via a user-ID filter. While physical structures are amortized, any single user's memories end up interleaved with other users' records across the collection's segments.

Recent vector-database releases introduce multitenancy optimizations for the shared layout. Qdrant supports designating the user-ID field as a \emph{tenant index}, prompting its optimizer to co-locate each tenant's points within a segment, build per-tenant HNSW graphs, and offload heavy tenants to dedicated shards~\cite{qdrantmultitenancy}. Milvus hashes a \emph{partition key} into a fixed set of physical partitions shared across tenants~\cite{milvusmultitenancy}. These mechanisms speed up locating a user's points inside a given segment or partition, yet a user's data still scatter across segments or partitions whose count and size are governed by the entire collection.

Both organizations work well at moderate user counts, but RedNote's production service must manage memories for 400 million users. Throughout this paper, we denote the total number of users by $C$, the total number of memories by $N$, and the number of memories owned by user $u$ by $m_u$. Section~\ref{sec:motivation} quantifies how storage and retrieval costs scale with $C$ and collection size under both organizations.

\begin{figure*}[t]
  \centering
  \includegraphics[width=\textwidth]{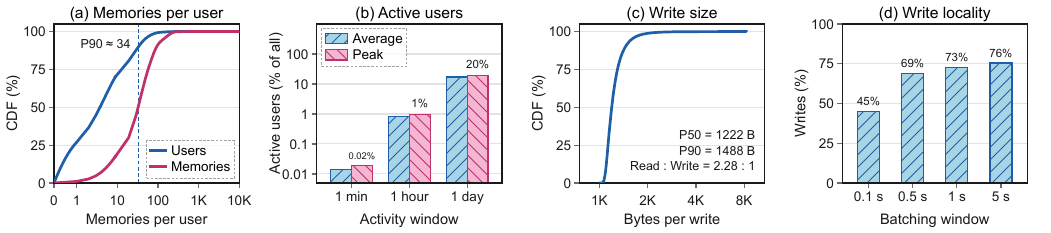}
  \caption{Production workload characteristics. (a) Distributions of users and
  memories by per-user memory count. (b) Active-user ratios over different time
  windows. (c) Write-size distribution. (d) Writes that can be coalesced by
  user within different time windows.}
  \label{fig:workload}
\end{figure*}

\section{Production Workload and Design Goals}
\label{sec:motivation}

\subsection{Production Memory Workload}
\label{sec:workload}

To understand the storage demands of production-scale user memory, we characterize RedNote's 400-million-user memory service using two complementary datasets: a full snapshot of historical memories across all users, and a seven-day trace of production requests. The snapshot captures how memories distribute across the user base, while the trace reveals activity patterns, write sizes, and temporal locality. Figure~\ref{fig:workload} highlights four properties that directly constrain the storage engine.

\paragraph{Heavy skew across user sizes.}
Although the 90th-percentile user holds only 34 memories, the user and memory cumulative distribution functions (CDFs) in Figure~\ref{fig:workload}a diverge sharply: the top 7.5\% of users account for 41.5\% of all stored memories, with a handful of users holding more than 1,000 records. A storage engine cannot optimize for only one end of this distribution:
per-user storage must incur minimal fixed overhead for the vast majority of small users while still supporting efficient retrieval for users with thousands of memories.

\paragraph{Sparse user activity.}
Only 0.02\%, 1\%, and 20\% of users are active within one-minute, one-hour, and one-day windows, respectively (Figure~\ref{fig:workload}b). Because the active working set is orders of magnitude smaller than the total population, keeping per-user state pinned in memory for idle users wastes DRAM at scale.

\paragraph{Read-dominated traffic with small writes.}
Production traffic averages 2.28 reads per write, and writes are small---1,222\,B at the median and 1,488\,B at the 90th percentile (Figure~\ref{fig:workload}c). While retrieval latency governs user-perceived performance, the aggregate write stream is high enough that paying a synchronous log flush and index update for every kilobyte-sized write is unsustainable.

\paragraph{Short-term write locality.}
We count a write as coalescible if the same user issues another write within a given time window. As Figure~\ref{fig:workload}d shows, the coalescible fraction jumps from 45\% at 0.1\,s to 73\% at 1\,s, then plateaus at 76\% by 5\,s. A sub-second, per-user batching window captures nearly all coalescing opportunities while keeping update visibility tight.

\subsection{Limitations of Existing Organizations}
\label{sec:existing-limitations}

\subsubsection{Experimental Setup}
\label{sec:motivation-setup}

We evaluate both the isolated and shared organizations on Qdrant~\cite{qdrant} using datasets sampled from the production distributions above, running on the same hardware as our main evaluation (\S\ref{sec:eval}). For the shared organization, we apply Qdrant's recommended multi-tenancy configuration~\cite{qdrantmultitenancy}: enabling \texttt{is\_tenant=true} on the user-ID payload index, disabling the global HNSW graph (\texttt{m=0}), and building per-tenant graphs via \texttt{payload\_m}.

\subsubsection{Isolated Collections Incur Excessive Per-User Overhead}
\label{sec:isolated}

\begin{figure}[t]
  \centering
  \includegraphics[width=\columnwidth]{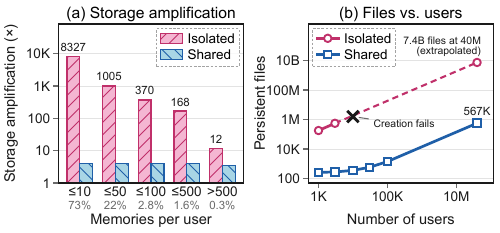}
  \caption{Costs of the isolated and shared organizations in Qdrant.
  (a) Storage amplification across per-user memory-count buckets.
  (b) Persistent-file count as the number of users grows.}
  \label{fig:isolation}
  \vspace{-3ex}
\end{figure}

Figure~\ref{fig:isolation}a reports storage amplification, defined as the ratio of on-disk footprint to logical memory size. For users with at most 10 memories---who make up 73\% of the user base---the isolated organization amplifies storage by 8,327$\times$. Amplification remains 1,005$\times$ for users with up to 50 memories and only drops to 12$\times$ beyond 500 memories. In contrast, the shared organization stays within single-digit amplification across all buckets, confirming that the isolated overhead stems from fixed per-collection machinery rather than the memory records themselves.

File-system metadata poses an even sharper cliff (Figure~\ref{fig:isolation}b). On a single node, collection creation fails at roughly 10,000 users due to file-descriptor exhaustion. Extrapolating the observed per-user file count linearly to 40 million users yields 7.4 billion persistent files, versus 567,000 files measured under the shared organization at the same scale.

\subsubsection{Shared Collections Compromise Query Efficiency}
\label{sec:shared}

\begin{figure}[t]
  \centering
  \includegraphics[width=\columnwidth]{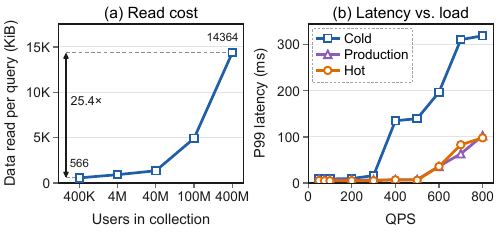}
  \caption{Query costs of the shared organization. (a) Average bytes read per
  query as the shared collection grows. (b) P99 latency under hot, cold, and
  production request streams.}
  \label{fig:shared}
\end{figure}

To isolate the impact of corpus scale, we build five nested shared collections ranging from 400 thousand to 400 million users. We sample 1,000 target users from a fixed 400,000-user subset following the production size distribution and replay production queries for each user, retrieving the top-20 memories. Although the target users and their memories remain identical across all seven collections, average bytes read per query climb by 25.4$\times$, from 566\,KiB at 400 thousand users to 14,364\,KiB at 400 million users (Figure~\ref{fig:shared}a). Even with Qdrant's multi-tenancy optimizations enabled, query I/O grows with the global collection rather than staying bounded by the queried user.

Figure~\ref{fig:shared}b shows the resulting tail latency on the 40-million-user collection under three request streams: \emph{hot} traffic repeatedly queries prewarmed users, \emph{cold} traffic hits an uncached user on every request, and \emph{production} traffic replays the user sequence
from production logs. At low to moderate load (50--500\,QPS), caching keeps hot and production P99 latency within 4.7--6.8\,ms and 6.7--7.8\,ms, respectively, whereas cold queries expose the underlying I/O amplification, climbing from 16.1\,ms at 300\,QPS to 135.0\,ms at 400\,QPS (19.8$\times$ the hot latency). Beyond 500\,QPS, however, all three streams cross a saturation knee and rise steeply, reaching 98--319\,ms at 800\,QPS. Even when target users are cached, per-query fan-out across the shared collection's segments degrades tail latency once concurrency increases.

Taken together, isolation preserves user-local I/O at an untenable per-user resource cost, whereas sharing amortizes storage structures but leaks global corpus scale into every query.

\subsection{Design Goals}
\label{sec:principles}

These measurements motivate three design goals for a production user-memory storage engine:
\begin{itemize}[leftmargin=*]
  \item \textbf{G1: Disk footprint follows data.} On-disk storage should scale with $N$ plus a fixed per-user overhead of only a few tens of bytes, eliminating the per-collection amplification of the isolated layout (\S\ref{sec:isolated}).
  \item \textbf{G2: Resident memory follows activity.} Beyond that minimal per-user overhead, host-DRAM consumption should scale with the active working set rather than growing with all $C$ users through resident metadata or file handles.
  \item \textbf{G3: Query work follows the queried user.} Pages read and entries scanned by a query should depend strictly on $m_u$, the query, and $k$, remaining independent of $C$, $N$, and other users' data (\S\ref{sec:shared}).
\end{itemize}

Two properties of our production workload (\S\ref{sec:workload}) further shape the design: per-user structures must stay compact for small users while still supporting heavy users with thousands of memories, and short-term write locality makes the user, rather than the individual record, the natural unit of update. \sys meets these goals through the \emph{virtual collection} abstraction, decoupling logical per-user isolation from physically shared storage (\S\ref{sec:design}).

\section{\sys Design}
\label{sec:design}
In this section, we present \sys, a storage engine designed for
production-scale LLM user memory that satisfies the design goals in
\S\ref{sec:principles}.
\subsection{Overview}
\label{sec:overview}

\paragraph{Virtual collections.}
\sys gives each user a \emph{virtual collection}. Logically, a virtual collection behaves like a private collection: it owns a dense space of memory identifiers (\emph{ordinals}), isolated keyword statistics, and independent access paths. Physically, it allocates no per-user files or segments and maintains no corpus-wide index. Instead, \sys separates collection naming, per-user structure, and physical placement across three levels: a compact resident \emph{user directory} (32\,B per user) locates every collection, an on-disk \emph{block map} describes each collection's internal layout, and a unified \emph{block space} packs all users' blocks into shared pages.

The abstraction mirrors virtual memory. Ordinals act as virtual addresses, each user's block map serves as a page table translating ordinals to block locations, and shared pages form the physical frames. This mapping directly
addresses our design goals (\S\ref{sec:principles}). Sub-page block allocation ties disk footprint to stored data (G1). On-demand paging of block maps ensures that, beyond a 32-byte directory entry per user, resident DRAM scales with active users (G2). Reading only through the target user's block
map keeps query work proportional to that user's data (G3).

\begin{figure}[t]
  \centering
  \includegraphics[width=\columnwidth]{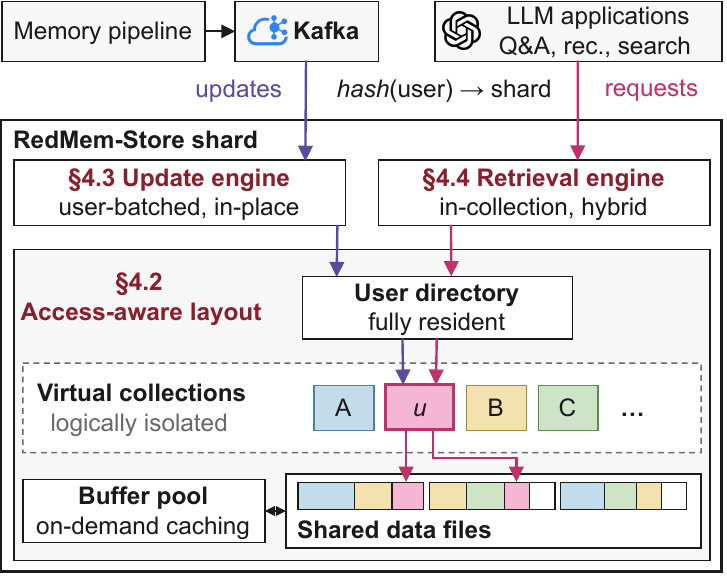}
  \caption{Architecture of \sys.}
  \label{fig:arch}
  \vspace{-3ex}
\end{figure}

\begin{figure*}[t]
  \centering
  \includegraphics[width=\textwidth]{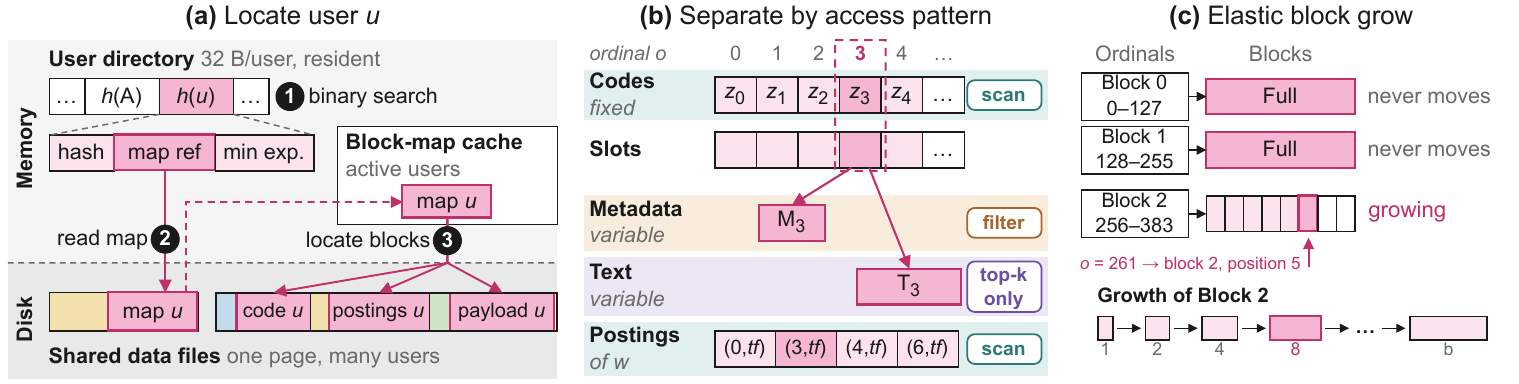}
  \caption{Access-aware virtual collection layout: (a) locating a collection
  through the resident user directory and the user's block map, (b) storing a
  user's memories in ordinal-aligned blocks separated by access type, and
  (c) growing code blocks and packing them into shared pages.}
  \label{fig:vcoll}
\end{figure*}

\paragraph{Architecture.}
Figure~\ref{fig:arch} shows the three core components of a shard. The \emph{access-aware layout} (\S\ref{sec:storage}) organizes virtual collections across a small, fixed set of shared data files through the user directory, block maps, and a buffer pool. The \emph{update engine} (\S\ref{sec:update}) consumes writes published by RedNote's memory pipeline to Kafka~\cite{kreps2011kafka}, coalesces them by user, and applies each user group in place. The \emph{retrieval engine} (\S\ref{sec:retrieval}) executes hybrid search within the queried user's collection. Periodically, an offline pipeline builds and publishes an immutable \emph{base} for each shard. A node loads the new base and replays the Kafka stream from the base build timestamp
to catch up.
\subsection{Access-Aware Virtual Collection Layout}
\label{sec:storage}

A virtual collection pairs an on-disk block map with data blocks partitioned
by access type (Figure~\ref{fig:vcoll}): quantized vector codes are packed
into contiguous code blocks, each term's postings form a compact array, and
metadata and raw text reside in per-memory blocks. All blocks are allocated
from a unified block space within the shard. We next describe how a request
locates a collection, how a collection stores and grows its memories, and how
collections share pages.

\paragraph{Locating a user at constant per-user resident cost.}
Keeping per-user state pinned in DRAM wastes memory at scale, since only
0.02\% of users are active in a typical minute window (Figure~\ref{fig:workload}b). \sys splits a user's state into two parts (Figure~\ref{fig:vcoll}a). The resident \emph{user directory} maintains an array of 32-byte entries sorted by user hash, each recording only the location of the user's block map and the earliest expiration timestamp among that user's memories. Everything else needed to locate a user's blocks remains in the on-disk \emph{block map} until accessed. A request runs a binary search in the directory, fetches the block map on a cache miss, and follows its pointers to the required blocks. Across 400 million users, the resident directory occupies $32C$ bytes (about 11.9\,GiB cluster-wide), while locating any user takes one in-memory search and at most one block-map read, with I/O independent of $C$ and $N$.

Decoded block maps are cached in memory under CLOCK replacement. On eviction,
clean maps are dropped, whereas dirty maps spill to a temporary auxiliary file
to reclaim DRAM and are discarded upon crash recovery. Once evicted, an idle
user consumes only its 32-byte directory entry. As in the working-set model of
virtual memory~\cite{denning1968workingset}, resident memory beyond the
directory scales with recent activity rather than the total user population.

\paragraph{Reading only what each stage needs.}
Hybrid retrieval executes stages with different data footprints: dense search
scans vector codes, sparse search reads postings for query terms, filtering
inspects selected metadata fields, and materialization fetches text for the
top-$k$ results. Storing memories as whole records would force every stage to
read unneeded fields. \sys instead assigns each memory a dense \emph{ordinal}
in $[0, m_u)$, recycling deleted ordinals, and separates fields by access type
(Figure~\ref{fig:vcoll}b). Fixed-size vector codes occupy ordinal-ordered code
blocks of up to $b$ entries, placing ordinal $o$ at position $o \bmod b$ of
block $\lfloor o/b \rfloor$ so the block map tracks only code blocks rather
than per-memory pointers. Variable-size metadata and text live in per-memory
blocks reached via each memory's \emph{slot} in the block map. Postings are
indexed by term, with each term's postings stored as a contiguous array of
$(\mathit{ordinal},\mathit{tf})$ pairs. Applying columnar
separation~\cite{ailamaki2001pax,abadi2007materialization} inside a single
user's collection lets dense search stream over contiguous codes without
per-record indirection and sparse search read only query-term arrays. Metadata
is decoded only for ordinals reached by a retrieval path, and text only for
the final results. Shared ordinals turn cross-stage combination into array
lookups (\S\ref{sec:retrieval}), and per-user scoping removes both per-record
user IDs and user-ID payload indexes.

\paragraph{Growing each structure independently.}
Collections evolve continuously across orders of magnitude in size
(\S\ref{sec:workload}). Pre-allocating space wastes storage on small users,
whereas rewriting a collection on every insert copies existing data repeatedly.
\sys grows each block type independently (Figure~\ref{fig:vcoll}c). A code
block starts with room for one code and doubles upon filling, up to a cap of
$b$ codes. Expansion copies only the growing block while full blocks
remain stationary, bounding amortized copy overhead to one code per insert and
unused space to at most one partial block per user. Updates overwrite existing
codes in place. When a term's posting array fills, \sys reallocates it to a
larger block; within an existing block, insertions, updates, and deletions
execute in place via an in-memory position index. Metadata and text blocks are
immutable: an update writes a new block and frees the old one without shifting
other memories. Collections exceeding $T$ memories add a Parent--Leaf tier
over their codes, placing leaf partitions in dedicated code blocks and parent
centroids in the block map. Ordinals remain invariant across this transition,
so posting lists, metadata slots, and ordinal-based rank fusion stay unchanged.

\paragraph{Packing collections onto shared pages.}
All collections in a shard allocate from a unified block space backed by a few
data files, using size classes and free lists to keep internal fragmentation
low even for blocks of a few dozen bytes. Storage scales with actual data,
small collections share pages, and file count stays independent of $C$. To
keep a user's blocks close together, the offline build ingests input grouped
by user and finishes with a \emph{compact export} that copies the live blocks
of consecutive users in order while reclaiming space freed by growth and
deletions. Between builds, in-place updates preserve this grouping except for
blocks that outgrow their capacity, which the next base build regroups. By
contrast, a shared vector collection scatters a user's points across segments
in insertion order. Because many small blocks share a page, the buffer pool
caches and flushes fixed-size pages rather than individual blocks.

\subsection{User-Batched In-Place Updates}
\label{sec:update}

\begin{figure}[t]
  \centering
  \includegraphics[width=\columnwidth]{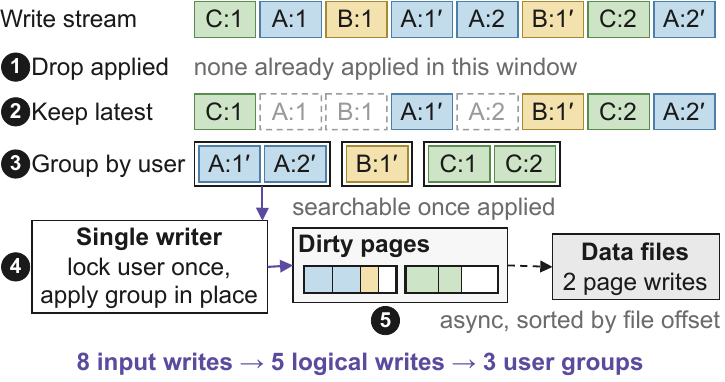}
  \caption{User-batched in-place updates.}
  \label{fig:update}
\end{figure}

Because memory writes are small, continuous, and clustered by user within about a second (Figure~\ref{fig:workload}c--d), \sys makes the user the unit of update, coalescing writes by user and applying each user group atomically in place (Figure~\ref{fig:update}).

\paragraph{Batch formation.}
The update engine seals a batch once a time, message-count, or byte limit is reached (\S\ref{sec:impl}), while continuing to accept incoming writes into the next buffer. Within each batch, it discards messages whose Kafka offsets were already applied, retains only the latest operation per memory, and groups the remaining operations by user. Because each operation replaces or deletes an entire memory, keeping only the latest one preserves the final state while collapsing redundant update and delete sequences. In production, a 500\,ms batch window coalesces 69\% of incoming writes with another write for the same user.

\paragraph{Pipelined per-user application.}
Applying a user group can touch uncached pages, such as pages holding old and
new term postings. Loader threads run a few users ahead of the writer,
decoding block maps and prefetching the pages each group will modify.
Prefetching is bounded both per user and as a fraction of the buffer pool
(\S\ref{sec:impl}), limiting the fraction of the pool that ingestion can
displace. A single writer then acquires each user's lock once and
applies the entire group in place: reusing free ordinals, overwriting
fixed-size codes in their slots, updating posting lists in $O(1)$ time via an
in-memory position index along with user-scoped keyword statistics, and
freeing empty blocks. For Parent--Leaf collections, the writer routes new codes to their nearest leaf and performs local leaf splits or merges as leaf sizes drift, following SPFresh~\cite{xu2023spfresh}. Concurrent queries see either all or none of a user group, while queries for other users proceed
independently. Kafka offsets advance only after every group in the batch
succeeds; if any group fails, the node halts consumption, reloads the base,
and replays the Kafka stream so no update is skipped.

\paragraph{Write-back and flow control.}
Modified pages stay dirty in the buffer pool and are flushed asynchronously
off the update path. Periodically, and whenever the dirty-page count exceeds a
low watermark (\S\ref{sec:impl}), a background task copies dirty pages with
their version numbers, sorts them by file offset, and writes them in
sequential batches. Pages modified during a flush remain dirty for the next
round. Because each page is written at most once per round regardless of how
many updates it absorbed, physical write volume scales with distinct dirty
pages rather than raw writes. When the dirty-page count exceeds a high
watermark, the writer flushes pages directly before applying further groups,
throttling ingestion before the pool fills.

\paragraph{Log-delegated durability.}
Because Kafka already stores every update durably in order, \sys keeps no
local write-ahead log (WAL) and takes no runtime checkpoints. Durable state
consists of the latest base plus the Kafka stream after the base build
timestamp. Each node modifies a writable copy of the base and marks it dirty
upon the first write. On clean shutdown, it writes back all dirty pages and
persists the directory alongside the applied offset of every Kafka partition.
After a crash, the node reloads the clean base and replays the Kafka stream.
During replay, the node skips already-applied records using per-memory write
positions and per-partition offsets, while routing all updates for a memory
through the same partition preserves ordering and avoids deletion tombstones.
Updates become visible as soon as the writer applies them to the buffer pool,
while Kafka guarantees persistence, avoiding the per-write synchronous flush
of a local WAL. The trade-off is a recovery time proportional to the replay
window, which is bounded by the base-build period, following the same design
point as changelog-backed stream processors~\cite{noghabi2017samza}.


\begin{figure}[t]
  \centering
  \includegraphics[width=\columnwidth]{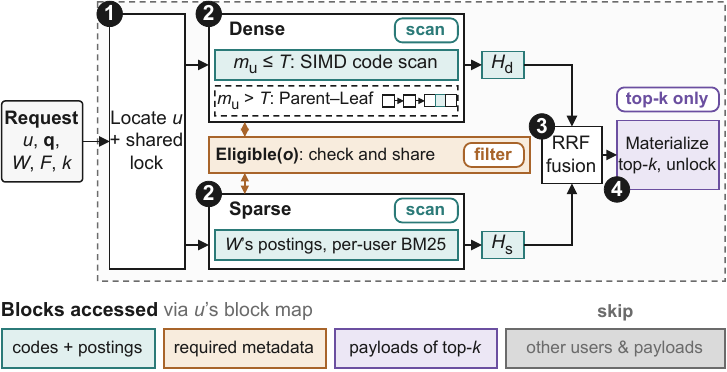}
  \caption{In-collection hybrid retrieval.}
  \label{fig:retrieval}
\end{figure}

\subsection{In-Collection Hybrid Retrieval}
\label{sec:retrieval}

Because most virtual collections are small and partitioned by access type,
scanning a user's compact structures is cheaper than traversing separate
vector and secondary indexes. \sys evaluates metadata filters, quantized
vector similarity, and keyword relevance within the queried collection
(Figure~\ref{fig:retrieval}). Algorithm~\ref{alg:retrieval} outlines the
execution path, where $k_c$ is the per-path candidate budget, $E$ and $A$ are
ordinal-indexed eligibility and BM25 arrays, and $H_d$ and $H_s$ are
size-$k_c$ candidate heaps.

\begin{algorithm}[t]
\caption{In-collection hybrid retrieval.}
\label{alg:retrieval}
\small
\begin{algorithmic}[1]
\Require user $u$, query vector $\mathbf{q}$, query terms $W$, filter $F$,
  sizes $k \le k_c$
\State lock $u$ shared; $\mathcal{C} \gets$ \Call{Locate}{$u$};
  $t \gets$ now\label{alg:locate}
\State $E[\cdot] \gets \bot$; $A[\cdot] \gets 0$ \Comment{$E$ is shared by both paths}\label{alg:elig-init}
\Function{Eligible}{$o$}\label{alg:ok-begin}
  \State \textbf{if} $E[o] = \bot$ \textbf{then}
    $E[o] \gets \mathit{live}(o) \land \mathit{exp}(o) > t \land F(o)$
  \State \Return $E[o]$\label{alg:ok-end}
\EndFunction
\If{$m_u \le T$}\label{alg:dense-begin}
  \State $\tilde{\mathbf{q}} \gets R\,\mathbf{q}$ \Comment{rotate once}
  \ForAll{code block $B$ of $\mathcal{C}$, ordinal $o \in B$ with \Call{Eligible}{$o$}}
    \State $H_d.\Call{Push}{o,\ s_o \langle \mathbf{z}_o, \tilde{\mathbf{q}} \rangle}$
      \Comment{SIMD scan}\label{alg:scan}
  \EndFor
\Else
  \State $H_d \gets \Call{LeafSearch}{\mathcal{C}, \mathbf{q}, k_c, \textsc{Eligible}}$
    \Comment{Parent--Leaf}\label{alg:index}
\EndIf
\ForAll{$w \in W$, $(o, \mathit{tf}) \in \mathcal{C}.\mathit{postings}[w]$ with \Call{Eligible}{$o$}}\label{alg:kw-begin}
  \State $A[o] \mathrel{+}= \mathrm{BM25}(\mathit{tf}, \ell_o;\ \mathcal{C}.\mathit{stats})$
    \Comment{user-scoped}
\EndFor
\State $H_s \gets$ top-$k_c$ ordinals with $A[o] > 0$\label{alg:kw-end}
\State $\mathcal{R} \gets$ top-$k$ of $H_d \cup H_s$ by
  $\sum_{p \in \{d,s\}:\,o \in H_p} 1/(k_{\mathrm{rrf}} + \mathit{rank}_p(o))$\label{alg:fuse}
\State $\mathcal{M} \gets \Call{Materialize}{\mathcal{R}}$; unlock $u$; \Return $\mathcal{M}$
  \Comment{texts of $\mathcal{R}$ only}\label{alg:mat}
\end{algorithmic}
\end{algorithm}

\paragraph{Ordinals as join keys.}
Every intermediate structure in a request operates directly on ordinals: the
eligibility array $E$ and BM25 accumulator $A$ are indexed by ordinal, while
the candidate heaps $H_d$ and $H_s$ store bare ordinals rather than
materialized records. Because both dense and sparse paths emit ordinals, rank
fusion reduces to lightweight ordinal-based merging akin to late
materialization in column stores~\cite{abadi2007materialization}. No
identifier translation or cross-index hash join is needed, and \sys fetches
text payloads and requested fields only for the final $k$ results
(line~\ref{alg:mat}) before releasing the shared user lock. Sizing $E$ and $A$
by $m_u$ keeps per-request scratch memory small for typical users.

\paragraph{Lazy, shared eligibility checks.}
Eligibility combines liveness, TTL expiration, and the metadata filter $F$,
and is cached across both paths in $E$
(lines~\ref{alg:elig-init}--\ref{alg:ok-end}). An ordinal is evaluated at most
once per request, unreached ordinals are never checked, and only the metadata
fields referenced by $F$ are decoded. Unlike filtered search over corpus-wide
proximity graphs, where selective filters can disconnect the graph or force
per-segment fallback plans~\cite{patel2024acorn,gollapudi2023filtereddiskann}
(\S\ref{sec:bg-vdb}), filtering inside a virtual collection skips ineligible
ordinals during scanning or posting traversal, avoiding both traversal
degradation and recall loss.

\paragraph{Scoring and fusion.}
Embeddings are encoded as 4-bit RaBitQ
codes~\cite{gao2024rabitq,gao2025extrabitq}. Because RaBitQ requires no
trained codebook, each memory is quantized independently of other users' data.
In Algorithm~\ref{alg:retrieval}, $\mathbf{z}_o$ and $s_o$ denote the
quantized code and scaling factor of ordinal $o$, and $R$ is the query
rotation matrix. For collections with at most $T$ memories, \sys rotates the
query once and runs a SIMD scan over their code blocks, reading one or two
blocks for a typical user without traversal-induced recall loss
(lines~\ref{alg:dense-begin}--\ref{alg:scan}). For larger collections, it
scans only the leaves selected by parent centroids (line~\ref{alg:index}).
The sparse path computes BM25~\cite{robertson2009bm25} using document
frequencies and average document length maintained for user $u$, isolating term
weights from other users' vocabularies
(lines~\ref{alg:kw-begin}--\ref{alg:kw-end}). Reciprocal rank
fusion~\cite{cormack2009rrf} with constant $k_{\mathrm{rrf}}$ merges the two
candidate lists (line~\ref{alg:fuse}).

\subsection{Implementation}
\label{sec:impl}

\sys is implemented in roughly 7K lines of C++ as a native index type inside
RedNote's in-house retrieval service, reusing its routing, RPC, and
version-management infrastructure. Each shard's block space spans at most 16
data files, with 16-byte aligned allocations rounded to 16\,B multiples up to
256\,B and powers of two above. Each full code block holds $b = 128$ 4-bit
1,024-dimensional codes (64\,KiB), and collections exceeding $T = 5,000$
memories enable the Parent--Leaf tier. The buffer pool manages 64\,KiB pages
under CLOCK replacement, reading data files with direct I/O through
\texttt{io\_uring}~\cite{axboe2019iouring} and merging concurrent misses on
the same page into a single read. We avoid \texttt{mmap} and the kernel page
cache because their readahead and eviction policies suit this layout
poorly~\cite{crotty2022mmap}. Distance scoring provides AVX2 and scalar
kernels, while per-user reader-writer locks and buffer-pool page pins allow
concurrent reads and updates across different users sharing a page. In
production, the update engine closes a batch after 500\,ms, 256 messages, or
4\,MiB, whichever comes first; loader threads prefetch up to four users ahead,
capped at 16 pages per user and one-eighth of the buffer pool.

\section{Evaluation}
\label{sec:eval}

We conduct a comprehensive experimental evaluation of \sys to investigate the following core aspects:
\begin{itemize}[leftmargin=1.5em]
  \item How do the query latency and write throughput of \sys scale compared
    with Qdrant and Milvus as the number of users grows from 400 thousand to
    400 million? (\S\ref{sec:eval-scale})
  \item How does \sys's resource consumption (read I/O, peak memory, disk
    footprint, and build time) compare with the baselines across scales?
    (\S\ref{sec:eval-resource})
  \item How does \sys's tail latency behave under increasing concurrent load
    across hot, cold, and production workloads? (\S\ref{sec:eval-load})
  \item Does replacing the storage engine with \sys preserve the end-to-end
    answer quality of an LLM memory system? (\S\ref{sec:eval-qa})
  \item How much do the access-aware layout and user-batched updates each
    contribute to read and write performance? (\S\ref{sec:eval-ablation})
\end{itemize}

\subsection{Experimental Setup}
\label{sec:eval-setup}

\paragraph{Testbed.} 
All experiments run on a server equipped with a 96-core AMD EPYC 9654 (Genoa) processor at 2.4\,GHz, 1,152\,GB of DDR5 memory (12$\times$96\,GB), and 12$\times$3.84\,TB SSDs, running Anolis OS 8.9
(Linux kernel 5.10.134).

\begin{figure*}[t]
  \centering
  \includegraphics[width=\textwidth]{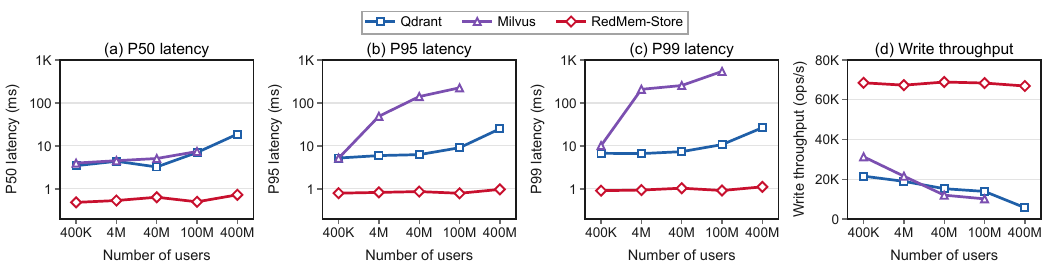}
  \caption{Query latency of cold queries and saturated write throughput as
  the number of users grows.}
  \label{fig:eval-overall}
\end{figure*}

\paragraph{Compared systems.}
We compare \sys against Qdrant~\cite{qdrant} and Milvus~\cite{wang2021milvus}, with each system storing all users in a single shard on this machine. Both baselines use their recommended multi-tenant configurations: Qdrant marks the user ID as a tenant index (\texttt{is\_tenant=true}) and builds per-tenant HNSW graphs (\texttt{m=0},
\texttt{payload\_m=16})~\cite{qdrantmultitenancy}, while Milvus designates the user ID as its partition key~\cite{milvusmultitenancy}. All three systems deploy the same RaBitQ technique to quantize vector embeddings.

\paragraph{Datasets.}
We evaluate on production memories from RedNote users. The largest dataset
holds the memories of 400 million users, and smaller datasets of 400 thousand,
4 million, 40 million, and 100 million users are drawn from the same production corpus. Each memory record contains raw text, metadata, and a 1,024-dimensional embedding. The total number of memories across all users is approximately 3.6 billion.

\paragraph{Workloads and metrics.}
Every query retrieves the top-20 memories for a single user. Query users are
sampled in proportion to their memory count, and we report the P50, P95, and P99 latency over all sampled queries. We evaluate three access patterns: \emph{cold} queries always target uncached users, \emph{hot} queries repeatedly access prewarmed users,
and \emph{production} queries replay the user sequence from RedNote's production logs to preserve real temporal locality. Concurrent queries are issued only in \S\ref{sec:eval-load} and all other latency measurements use
single-threaded cold queries. Write throughput is measured by scaling client threads until throughput saturates. For resource efficiency, we track bytes read from storage per query, peak serving-process memory during queries, on-disk size of data and indexes, and index build time from an empty database.

\paragraph{Answer quality.}
To assess end-to-end quality, we run the state-of-the-art Mem0 memory framework~\cite{chhikara2025mem0} while swapping only its storage backend. In our experiment, Mem0 extracts memories using Qwen3-30B-A3B-Instruct-2507~\cite{yang2025qwen3},
answers each question from the top-20 retrieved memories, and the correctness of the answer is evaluated by CompassJudger-2-32B Instruct~\cite{zhang2025compassjudger2}.
We use standard benchmarks for evaluating memory system effectiveness, including LoCoMo~\cite{maharana2024locomo} and LongMemEval-S~\cite{wu2025longmemeval}.


\subsection{Performance Scaling}
\label{sec:eval-scale}

Figure~\ref{fig:eval-overall} reports cold-query latency and saturated write
throughput as the user population scales by three orders of magnitude, from
400 thousand to 400 million users. We omit Milvus at 400 million users because building its index at that scale exceeds ten days on our testbed (compared with 4 hours for \sys and 18 hours for Qdrant; \S\ref{sec:eval-resource}).

\paragraph{Cold-query latency.}
Across the entire user range, \sys keeps P50 latency between 0.49 and 0.73\,ms and P99 latency between 0.91 and 1.12\,ms. Two properties explain
this invariance. First, locating a user via binary search over the resident 32-byte user directory is purely in-memory and scales logarithmically with
$C$, adding negligible CPU time even at 400 million users. Second, once the user's block map is fetched, retrieval reads only the pages containing that
user's blocks, so the I/O path never depends on how many other users share the shard.

By contrast, both baselines incur substantially higher latency and degrade as
the collection grows. Qdrant's P50 latency spans 3.3--18.9\,ms
(5.1--25.8$\times$ slower than \sys), with P99 latency 7.0--23.8$\times$
higher. Milvus averages 8.0--14.8$\times$ slower than \sys and exhibits steep tail degradation: its P99 climbs from 10.2\,ms at 400 thousand users to
542\,ms at 100 million users, 590$\times$ that of \sys. Even with tenant indexes or partition keys enabled, a shared collection accumulates more
segments or partitions as $C$ increases. A single user's memories remain scattered across these global units, forcing each query to fan out across segments, probe per-segment metadata structures, and issue scattered disk reads that grow with the corpus rather than with the queried user.

\paragraph{Saturated write throughput.}
Write throughput shows an equally clear separation across scales. \sys sustains 66.9K--68.9K writes per second from 400 thousand to 400 million users. Because typical collections hold few memories and require no graph or clustered index, applying a user's coalesced batch reduces to overwriting
fixed-size code slots in place and appending or updating small posting and payload blocks (\S\ref{sec:update}). The cost of each write therefore depends only on the affected user and remains decoupled from $C$ and $N$. In the baselines, however, ingesting continuous small writes into a growing shared collection triggers heavier segment sealing, background compaction, and per-tenant or per-segment index maintenance. As a result, Qdrant's throughput declines from 21.6K to 5.7K writes per second and Milvus's falls from 31.3K to 10.1K, leaving \sys 3.2--11.7$\times$ faster than Qdrant and 2.2--6.8$\times$ faster than Milvus.

\subsection{Resource Efficiency}
\label{sec:eval-resource}

\begin{figure*}[t]
  \centering
  \includegraphics[width=\textwidth]{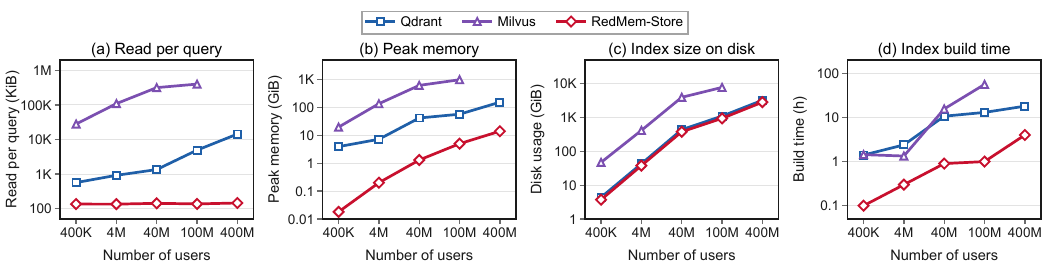}
  \caption{Resource usage as the number of users grows: bytes read per cold
  query, peak memory during queries, on-disk size, and index build time.}
  \label{fig:eval-resource}
\end{figure*}

\begin{figure*}[t]
  \centering
  \includegraphics[width=\textwidth]{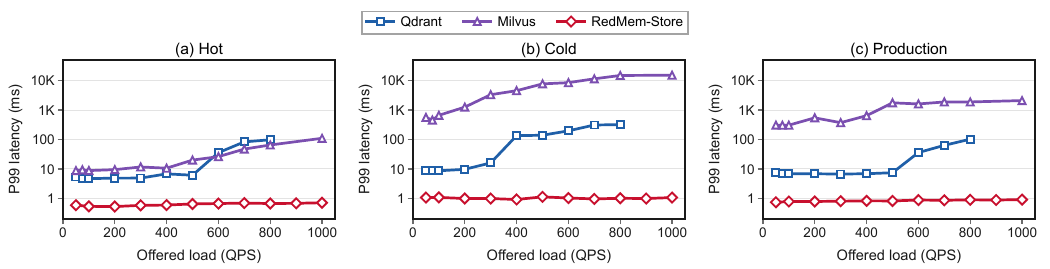}
  \caption{P99 latency under increasing offered load with 40 million users
  for the hot, cold, and production workloads.}
  \label{fig:eval-load}
\end{figure*}

Figure~\ref{fig:eval-resource} breaks down read I/O, peak memory, on-disk footprint, and index build time across scales, directly mapping each engine's
resource usage onto our design goals (\S\ref{sec:principles}).

\paragraph{Bytes read per query.}
\sys achieves the lowest per-query read volume among all systems, reading only 135--145\,KiB per cold query at every scale. Because compact export groups each user's live blocks onto shared pages and retrieval touches only the access paths required by the query, read volume tracks the queried user's data and satisfies scale independence. Both baselines violate this bound as the corpus expands.
In Qdrant, per-query read volume grows 25$\times$, from
567\,KiB at 400 thousand users to 14,364\,KiB at 400 million users (99$\times$ more bytes per query than \sys at the largest scale), because each query must read index files whose size grows with the number of users. Milvus reads 27\,MiB per query even at 400 thousand users and climbs to 392\,MiB at 100 million users, roughly 2,900$\times$ that of \sys.

\paragraph{Peak memory.}
Peak memory in \sys increases from 0.018\,GiB at 400 thousand users to
14\,GiB at 400 million users, tracking the fixed resident user directory. At
400 million users, the 32-byte directory entries consume 11.9\,GiB (85\% of
the process footprint), while the remaining 2.1\,GiB holds decoded block maps
for active users and cached buffer-pool pages. Because idle users drop back
to their 32-byte directory entries, memory beyond the directory follows
active traffic. Qdrant requires 153.5\,GiB at 400 million users
(11$\times$ more than \sys), and Milvus requires 615\,GiB already at 40
million users, as both systems maintain substantial per-segment index
metadata, payload index structures, and graph or partition state in memory.
Keeping per-user DRAM overhead this low directly translates to higher user
density per serving node in production.

\paragraph{Disk usage.}
The on-disk footprint of \sys scales linearly with stored memories, growing
from 3.7\,GiB at 400 thousand users to 2,833\,GiB at 400 million users. This
is 12--15\% smaller than Qdrant and 8--13$\times$ smaller than Milvus. Because
both \sys and Qdrant store quantized vectors alongside identical memory text
and amortize global files across users, their on-disk sizes remain close. The
modest reduction in \sys comes mainly from eliminating repeated per-record
user IDs and payload indexes. Compared with shared Qdrant, the primary efficiency gains of \sys therefore lie in memory consumption (11$\times$ lower) and query read I/O (99$\times$ lower), while still satisfying the ``footprint follows data'' principle without the storage amplification of isolated collections.

\paragraph{Index build time.}
\sys builds the 400-million-user base in 4 hours, compared with 18 hours for Qdrant, whereas Milvus cannot finish within a practical build window at this scale (requiring over ten days). At 100 million users, \sys builds 13$\times$ faster than Qdrant and 56$\times$ faster than Milvus. Two design choices drive this gap. First, typical collections need no graph or clustered vector index, avoiding costly neighbor search during ingestion. Second, 4-bit RaBitQ quantization requires no corpus-wide codebook training, allowing the offline builder to encode each user's memories independently and write live blocks sequentially through compact export. Beyond saving offline compute, faster base builds have a direct operational benefit: they allow shorter intervals between bases, which in turn bounds the Kafka replay window and crash-recovery time on serving nodes (\S\ref{sec:update}).


\subsection{Latency under Load}
\label{sec:eval-load}

Figure~\ref{fig:eval-load} evaluates P99 latency on the 40-million-user
dataset as offered load rises from 50 to 1,000 queries per second (QPS) across
the hot, cold, and production workloads. \sys holds P99 latency at or below
1.13\,ms across all three workloads and all tested QPS levels. At 1,000 QPS,
its P99 is 0.71\,ms for hot queries, 0.91\,ms for production queries, and
1.07\,ms for cold queries. Notably, a cold cache miss adds only about
0.4\,ms over a hot hit. Because each cold query reads only a small number of
pages via direct I/O with \texttt{io\_uring} and acquires a per-user shared
lock, concurrent requests neither saturate SSD bandwidth nor contend across
users, keeping tail latency flat up to peak offered load.

Both baselines hit sharp saturation cliffs as concurrency increases. Qdrant
keeps cold P99 below 10\,ms up to 200 QPS, but latency climbs to 135\,ms at
400 QPS and reaches 319\,ms at 800 QPS, where \sys is 316$\times$ faster.
Qdrant's hot and production curves remain below 8\,ms through 500 QPS before
exceeding 35\,ms from 600 QPS onward. Milvus answers hot queries in roughly
10\,ms at low load, yet its cold P99 already reaches 563\,ms at 50 QPS and
degrades to 15.2\,s at 1,000 QPS. Across all three engines, the production
curve tracks between the hot and cold bounds. While temporal locality in
production traffic helps reduce average latency, only 0.02\% of users are
active within any one-minute window (Figure~\ref{fig:workload}b), so cold
accesses remain frequent enough to dominate the 99th percentile. When each
cold miss triggers tens to hundreds of times more read I/O, queuing behind
cold disk reads degrades overall tail latency; \sys avoids this bottleneck by
making the cold path itself inexpensive.

\subsection{Answer Quality}
\label{sec:eval-qa}

\begin{figure}[t]
  \centering
  \includegraphics[width=\columnwidth]{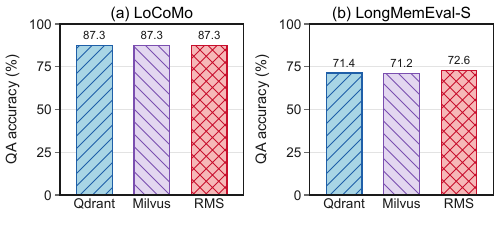}
  \caption{QA accuracy of Mem0 with different storage engines. RMS denotes RedMem-Store.}
  \label{fig:eval-qa}
\end{figure}

Figure~\ref{fig:eval-qa} verifies that replacing the storage backend preserves
the end-to-end answer quality of an LLM memory system. Running the same Mem0
pipeline, extraction model, and judge across all three engines yields approximately 87.3\% accuracy on LoCoMo regardless of the backend. On LongMemEval-S, Mem0 achieves
72.6\% accuracy with \sys, compared with 71.4\% with Qdrant and 71.2\% with
Milvus. These results confirm that dropping global or per-tenant graph indexes
for typical collections does not sacrifice retrieval effectiveness. Scanning
4-bit RaBitQ codes over every eligible memory in a user's virtual collection
avoids the traversal-induced recall loss that can occur when filtered graph
search navigates sparse per-tenant subgraphs. In addition, maintaining BM25
document-frequency and length statistics per user isolates keyword scoring
from other users' vocabularies during hybrid fusion, supplying the answering
LLM with context at least as relevant as that returned by the baselines.

\subsection{Ablation Study}
\label{sec:eval-ablation}

\begin{figure}[t]
  \centering
  \includegraphics[width=\columnwidth]{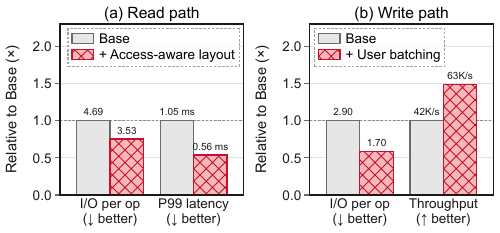}
  \caption{Effect of the access-aware layout on reads and of user batching
  on writes (40 million users, hot workload).}
  \label{fig:eval-ablation}
  \vspace{-3ex}
\end{figure}

Figure~\ref{fig:eval-ablation} isolates the individual contributions of our
layout and update mechanisms on the 40-million-user dataset under the hot
workload.

\paragraph{Access-aware layout.}
On the read path, we compare \sys against a variant that stores each memory
as a single monolithic record containing its vector code, metadata, and raw
text together. Under the monolithic layout, dense retrieval and sparse
retrieval must pull full records whenever they inspect a memory, inflating
page traffic with metadata fields and text strings that are never needed by
discarded candidates. Separating fields by access type (\S\ref{sec:storage})
reduces I/O operations per query by 25\% (from 4.69 to 3.53) and lowers P99
latency by 46\% (from 1.05 to 0.56\,ms). Because dense search streams over
contiguous code blocks, eligibility checks decode only referenced metadata
fields for reached ordinals, and text payloads are fetched only for the final
top-$k$ results, both I/O count and tail latency drop substantially.

\paragraph{User-batched updates.}
On the write path, we compare \sys against a baseline that applies every
incoming write individually in place without per-user batching. Applying
writes one by one forces the writer to acquire the user lock, locate and
decode the block map, and dirty shared pages separately for every small
kilobyte-sized write. Grouping writes by user within the batching window
(\S\ref{sec:update}) cuts I/O operations per write by 41\% (from 2.90 to
1.70) and raises saturated write throughput by 48\% (from 42.5K to 62.9K
writes per second). Two effects account for this improvement: collapsing
multiple writes to the same memory eliminates redundant storage updates
altogether, and applying the remaining writes for a user together amortizes
block-map loads, prefetching, page reads, and asynchronous page write-backs
across the entire user group.

\section{Deployment Experience and Lessons}
\label{sec:lessons}

\paragraph{From shared in-memory indexes to a dedicated engine.}
RedNote's memory service went through two earlier designs before \sys. Lacking
a mature storage backend for production-scale user memory, we initially stored
all memories in a single in-memory vector collection and answered each query
with a user-ID pre-filter. Once the corpus exceeded 10 million memories, query
latency climbed to several hundred milliseconds, leaving too little of the
request budget for agent planning and LLM inference. We then built a plugin
that mapped each user ID to the user's vector IDs, fetched those vectors, and
scored them directly. Latency dropped back to tens of milliseconds, but every
vector still had to remain resident in memory. As the corpus passed 10 billion
memories and more products onboarded, the cluster expanded to nodes that
exhausted DRAM while CPU utilization remained low. This imbalance between
memory and compute, even more than latency, drove us to build \sys.


\paragraph{Organize storage around the user, not the collection.}
Each query needs only a single user's few dozen memories, yet our first design
forced it to pay for the entire corpus, tying latency to total corpus size.
Per-user collections remove that scan overhead but multiply fixed metadata and
file overhead by hundreds of millions. Although multi-tenant features in
current vector databases narrow this gap, Qdrant still reads 25$\times$ more
bytes per query at 400 million users than at 400 thousand
(\S\ref{sec:eval-resource}).
Making the user the unit of physical organization and keeping per-user fixed
state to a few tens of bytes proved far more effective than tuning
collection-level structures.

\paragraph{For small per-user datasets, scanning beats indexing.}
Our plugin already demonstrated that a vector index is unnecessary once
retrieval is confined to one user: scoring that user's vectors exhaustively
cut latency from hundreds to tens of milliseconds. \sys takes this further by
packing each user's quantized codes contiguously, allowing a query to scan one
or two code blocks with SIMD instructions. Direct scanning preserves
end-to-end answer quality (\S\ref{sec:eval-qa}) and keeps index maintenance
off the write path. As a result, \sys builds its index several times faster
than the baselines and sustains flat write throughput as the user population
grows (\S\ref{sec:eval-scale}, \S\ref{sec:eval-resource}). Clustered indexing
is reserved for the small tail of users exceeding the scan threshold
(\S\ref{sec:retrieval}).

\paragraph{Since DRAM is expensive, optimizations on the cold path become crucial. }
In our plugin-based cluster, DRAM often exhausts while most CPU cores remain idle. Offloading memory data to SSD relieves this bottleneck only if cold accesses remain inexpensive. Because only a small fraction of users are active in any
given minute (Figure~\ref{fig:workload}b), most requests originate from uncached
users, and no affordable cache capacity changes that ratio. \sys bounds each cold
access to one lookup in the resident user directory and a few page reads.
Under concurrent load, a cold query in \sys adds only about 0.4\,ms over a hot
hit, whereas cold P99 latency reaches up to 23$\times$ the hot latency in
Qdrant and 428$\times$ in Milvus (\S\ref{sec:eval-load}).

\section{Related Work}
\label{sec:related}

\paragraph{Memory systems for LLM agents.}
An LLM memory system extracts salient facts from each interaction,
consolidates them with the user's existing memories, and retrieves the most
relevant ones into the prompt of later requests. Recent memory frameworks such as Mem0~\cite{chhikara2025mem0} perform extraction and consolidation with LLM calls that add, update, or delete memories, and MemOS~\cite{li2025memos} manages memories of different types and lifetimes as a system resource. OpenViking~\cite{openviking} manages an agent's memories, resources, and skills as a file system with a directory per user and loads context level by level, from directory summaries down to full content. These systems decide what to remember and how to supply it to the model, and they keep memories in general-purpose databases.  \sys is complementary to these systems: it replaces the storage layer beneath them and leaves the memory pipeline
unchanged.

\paragraph{Vector indexes and acceleration.}
Approximate nearest-neighbor (ANN) search over large vector corpora relies on
in-memory graph indexes such as HNSW~\cite{malkov2018hnsw} and quantization
methods such as PQ~\cite{jegou2011pq} and RaBitQ~\cite{gao2024rabitq}. To
scale beyond DRAM capacity, disk-based indexes such as
DiskANN~\cite{subramanya2019diskann} and SPANN~\cite{chen2021spann} pair
compact in-memory navigation structures with on-disk storage. Recent systems
also accelerate large-scale vector search across heterogeneous hardware~\cite{tian2025fusionanns}. All of these techniques
target a single large corpus and aim for fast search over millions to
billions of vectors. In contrast, production user memory consists of hundreds
of millions of small, independent datasets, where direct scanning is more
efficient than graph traversal. \sys defaults to SIMD scanning over 4-bit
RaBitQ codes for each user and builds a dedicated clustered index only for
large users whose collections exceed the scan threshold
(\S\ref{sec:retrieval}).

\paragraph{Vector databases and multi-tenancy.}
General-purpose open-source vector databases have introduced multi-tenant features to
serve many users.
Qdrant~\cite{qdrant} allows designating the user ID as a tenant index, which
co-locates each tenant's points within segments, builds per-tenant HNSW
graphs, and offloads heavy tenants to dedicated
shards~\cite{qdrantmultitenancy}. Milvus~\cite{wang2021milvus} hashes a
partition key into a fixed set of physical partitions shared across tenants so
that queries probe only the matching partition~\cite{milvusmultitenancy}.
Although these optimizations speed up locating a tenant's points inside a
segment or partition, storage layout, indexing, and search execution remain
tied to segment granularity. As the shared collection grows, a user's memories
still scatter across global segments, causing query I/O and memory overhead to
scale with the corpus (\S\ref{sec:eval}). \sys breaks this coupling by making
the per-user virtual collection the unit of physical organization, keeping both resident state and query work independent of the total user count and corpus size.

\section{Conclusion}
\label{sec:conclusion}


We presented \sys, a virtual-collection storage engine for production-scale
LLM user memory. Combining an access-aware layout, user-batched in-place
updates, and in-collection hybrid retrieval over physically shared storage,
\sys eliminates per-user fixed overhead while confining query work to the
queried user. Across workloads up to 400 million users, \sys sustains
1\,ms P99 latency, outperforms Qdrant and Milvus by up to 25.8$\times$ in
query latency and 11.7$\times$ in write throughput, and reduces production
hardware costs at RedNote by over 90\%.
\bibliographystyle{ACM-Reference-Format}
\bibliography{refs}

@misc{chhikara2025mem0,
  archiveprefix = {arXiv},
  author = {Chhikara, Prateek and Khant, Dev and Aryan, Saket and Singh, Taranjeet and Yadav, Deshraj},
  doi = {10.48550/arXiv.2504.19413},
  eprint = {2504.19413},
  publisher = {arXiv},
  title = {{Mem0}: Building Production-Ready {AI} Agents with Scalable Long-Term Memory},
  url = {https://arxiv.org/abs/2504.19413},
  year = {2025}
}

@misc{li2025memos,
  archiveprefix = {arXiv},
  author = {Li, Zhiyu and Xi, Chenyang and Li, Chunyu and Chen, Ding and Chen, Boyu and Song, Shichao and Niu, Simin and Wang, Hanyu and Yang, Jiawei and Tang, Chen and Yu, Qingchen and Zhao, Jihao and Wang, Yezhaohui and Liu, Peng and Lin, Zehao and Wang, Pengyuan and Huo, Jiahao and Chen, Tianyi and Chen, Kai and Li, Kehang and Tao, Zhen and Lai, Huayi and Wu, Hao and Tang, Bo and Wang, Zhengren and Fan, Zhaoxin and Zhang, Ningyu and Zhang, Linfeng and Yan, Junchi and Yang, Mingchuan and Xu, Tong and Xu, Wei and Chen, Huajun and Wang, Haofen and Yang, Hongkang and Zhang, Wentao and Xu, Zhi-Qin John and Chen, Siheng and Xiong, Feiyu},
  doi = {10.48550/arXiv.2507.03724},
  eprint = {2507.03724},
  publisher = {arXiv},
  title = {{MemOS}: A Memory {OS} for {AI} System},
  url = {https://arxiv.org/abs/2507.03724},
  year = {2025}
}

@misc{packer2023memgpt,
  archiveprefix = {arXiv},
  author = {Packer, Charles and Wooders, Sarah and Lin, Kevin and Fang, Vivian and Patil, Shishir G. and Stoica, Ion and Gonzalez, Joseph E.},
  doi = {10.48550/arXiv.2310.08560},
  eprint = {2310.08560},
  publisher = {arXiv},
  title = {{MemGPT}: Towards {LLMs} as Operating Systems},
  url = {https://arxiv.org/abs/2310.08560},
  year = {2023}
}

@misc{zhang2024survey,
  archiveprefix = {arXiv},
  author = {Zhang, Zeyu and Bo, Xiaohe and Ma, Chen and Li, Rui and Chen, Xu and Dai, Quanyu and Zhu, Jieming and Dong, Zhenhua and Wen, Ji-Rong},
  doi = {10.48550/arXiv.2404.13501},
  eprint = {2404.13501},
  publisher = {arXiv},
  title = {A Survey on the Memory Mechanism of Large Language Model based Agents},
  url = {https://arxiv.org/abs/2404.13501},
  year = {2024}
}

@misc{qdrant,
  author = {{Qdrant}},
  howpublished = {\url{https://qdrant.tech}},
  key = {Qdrant},
  note = {Accessed October 2026},
  title = {{Qdrant}: Vector Search Engine},
  year = {2026}
}

@inproceedings{wang2021milvus,
  author = {Wang, Jianguo and Yi, Xiaomeng and Guo, Rentong and Jin, Hai and Xu, Peng and Li, Shengjun and Wang, Xiangyu and Guo, Xiangzhou and Li, Chengming and Xu, Xiaohai and Yu, Kun and Yuan, Yuxing and Zou, Yinghao and Long, Jiquan and Cai, Yudong and Li, Zhenxiang and Zhang, Zhifeng and Mo, Yihua and Gu, Jun and Jiang, Ruiyi and Wei, Yi and Xie, Charles},
  booktitle = {Proceedings of the 2021 International Conference on Management of Data (SIGMOD)},
  doi = {10.1145/3448016.3457550},
  pages = {2614--2627},
  title = {{Milvus}: A Purpose-Built Vector Data Management System},
  url = {https://doi.org/10.1145/3448016.3457550},
  year = {2021}
}

@article{malkov2018hnsw,
  author = {Malkov, Yu A. and Yashunin, Dmitry A.},
  doi = {10.1109/TPAMI.2018.2889473},
  journal = {IEEE Transactions on Pattern Analysis and Machine Intelligence},
  number = {4},
  pages = {824--836},
  title = {Efficient and Robust Approximate Nearest Neighbor Search Using Hierarchical Navigable Small World Graphs},
  url = {https://doi.org/10.1109/TPAMI.2018.2889473},
  volume = {42},
  year = {2020}
}

@article{robertson2009bm25,
  title   = {The Probabilistic Relevance Framework: {BM25} and Beyond},
  author  = {Robertson, Stephen and Zaragoza, Hugo},
  journal = {Foundations and Trends in Information Retrieval},
  volume  = {3},
  number  = {4},
  pages   = {333--389},
  year    = {2009}
}

@inproceedings{gollapudi2023filtereddiskann,
  author = {Gollapudi, Siddharth and Karia, Neel and Sivashankar, Varun and Krishnaswamy, Ravishankar and Begwani, Nikit and Raz, Swapnil and Lin, Yiyong and Zhang, Yin and Mahapatro, Neelam and Srinivasan, Premkumar and Singh, Amit and Simhadri, Harsha Vardhan},
  title = {{Filtered-DiskANN}: Graph Algorithms for Approximate Nearest Neighbor Search with Filters},
  booktitle = {Proceedings of the ACM Web Conference 2023},
  pages = {3406--3416},
  year = {2023},
  publisher = {ACM},
  doi = {10.1145/3543507.3583552},
}

@article{jegou2011pq,
  author = {J{\'e}gou, Herv{\'e} and Douze, Matthijs and Schmid, Cordelia},
  title = {Product Quantization for Nearest Neighbor Search},
  journal = {IEEE Transactions on Pattern Analysis and Machine Intelligence},
  volume = {33},
  number = {1},
  pages = {117--128},
  year = {2011},
  doi = {10.1109/TPAMI.2010.57},
}

@article{patel2024acorn,
  author = {Patel, Liana and Kraft, Peter and Guestrin, Carlos and Zaharia, Matei},
  title = {{ACORN}: Performant and Predicate-Agnostic Search Over Vector Embeddings and Structured Data},
  journal = {Proceedings of the ACM on Management of Data},
  volume = {2},
  number = {3},
  pages = {1--27},
  year = {2024},
  doi = {10.1145/3654923},
}

@inproceedings{subramanya2019diskann,
  author = {Subramanya, Suhas Jayaram and Devvrit, Fnu and Kadekodi, Rohan and Krishnaswamy, Ravishankar and Simhadri, Harsha Vardhan},
  title = {{DiskANN}: Fast Accurate Billion-Point Nearest Neighbor Search on a Single Node},
  booktitle = {Advances in Neural Information Processing Systems},
  volume = {32},
  pages = {13748--13758},
  year = {2019},
}

@inproceedings{xu2023spfresh,
  author = {Xu, Yuming and Liang, Hengyu and Li, Jin and Xu, Shuotao and Chen, Qi and Zhang, Qianxi and Li, Cheng and Yang, Ziyue and Yang, Fan and Yang, Yuqing and Cheng, Peng and Yang, Mao},
  title = {{SPFresh}: Incremental In-Place Update for Billion-Scale Vector Search},
  booktitle = {Proceedings of the 29th Symposium on Operating Systems Principles},
  pages = {545--561},
  year = {2023},
  publisher = {ACM},
  doi = {10.1145/3600006.3613166},
}

@inproceedings{chen2021spann,
  author       = {Qi Chen and
                  Bing Zhao and
                  Haidong Wang and
                  Mingqin Li and
                  Chuanjie Liu and
                  Zengzhong Li and
                  Mao Yang and
                  Jingdong Wang},
  title        = {{SPANN:} Highly-efficient Billion-scale Approximate Nearest Neighbor
                  Search},
  booktitle    = {Proceedings of the International Conference on Neural Information Processing Systems (NeurIPS)},
  pages        = {5199--5212},
  year         = {2021}
}

@inproceedings {tian2025fusionanns,
author = {Bing Tian and Haikun Liu and Yuhang Tang and Shihai Xiao and Zhuohui Duan and Xiaofei Liao and Hai Jin and Xuecang Zhang and Junhua Zhu and Yu Zhang},
title = {Towards High-throughput and Low-latency Billion-scale Vector Search via {CPU/GPU} Collaborative Filtering and Re-ranking},
booktitle = {23rd USENIX Conference on File and Storage Technologies (FAST 25)},
year = {2025},
isbn = {978-1-939133-45-8},
address = {Santa Clara, CA},
pages = {171--185},
url = {https://www.usenix.org/conference/fast25/presentation/tian-bing},
publisher = {USENIX Association},
month = feb
}

@inproceedings{cormack2009rrf,
  author = {Cormack, Gordon V. and Clarke, Charles L. A. and B{\"u}ttcher, Stefan},
  booktitle = {Proceedings of the 32nd International ACM SIGIR Conference on Research and Development in Information Retrieval},
  doi = {10.1145/1571941.1572114},
  pages = {758--759},
  publisher = {Association for Computing Machinery},
  title = {Reciprocal Rank Fusion Outperforms {Condorcet} and Individual Rank Learning Methods},
  url = {https://doi.org/10.1145/1571941.1572114},
  year = {2009}
}

@article{gao2024rabitq,
  articleno = {167},
  author = {Gao, Jianyang and Long, Cheng},
  doi = {10.1145/3654970},
  journal = {Proceedings of the ACM on Management of Data},
  number = {3},
  numpages = {27},
  publisher = {Association for Computing Machinery},
  title = {{RaBitQ}: Quantizing High-Dimensional Vectors with a Theoretical Error Bound for Approximate Nearest Neighbor Search},
  url = {https://doi.org/10.1145/3654970},
  volume = {2},
  year = {2024}
}

@article{gao2025extrabitq,
  articleno = {202},
  author = {Gao, Jianyang and Gou, Yutong and Xu, Yuexuan and Yang, Yongyi and Long, Cheng and Wong, Raymond Chi-Wing},
  doi = {10.1145/3725413},
  journal = {Proceedings of the ACM on Management of Data},
  number = {3},
  numpages = {26},
  publisher = {Association for Computing Machinery},
  title = {Practical and Asymptotically Optimal Quantization of High-Dimensional Vectors in {Euclidean} Space for Approximate Nearest Neighbor Search},
  url = {https://doi.org/10.1145/3725413},
  volume = {3},
  year = {2025}
}

@inproceedings{kreps2011kafka,
  author    = {Kreps, Jay and Narkhede, Neha and Rao, Jun},
  title     = {Kafka: A Distributed Messaging System for Log Processing},
  booktitle = {Proceedings of the NetDB Workshop},
  year      = {2011}
}

@inproceedings{crotty2022mmap,
  author    = {Crotty, Andrew and Leis, Viktor and Pavlo, Andrew},
  title     = {Are You Sure You Want to Use {MMAP} in Your Database Management System?},
  booktitle = {Conference on Innovative Data Systems Research (CIDR)},
  year      = {2022}
}

@misc{axboe2019iouring,
  author       = {Axboe, Jens},
  title        = {Efficient {IO} with io\_uring},
  howpublished = {\url{https://kernel.dk/io_uring.pdf}},
  year         = {2019}
}

@inproceedings{aulbach2008multitenant,
  author    = {Aulbach, Stefan and Grust, Torsten and Jacobs, Dean and Kemper, Alfons and Rittinger, Jan},
  title     = {Multi-Tenant Databases for Software as a Service: Schema-Mapping Techniques},
  booktitle = {Proceedings of the 2008 ACM SIGMOD International Conference on Management of Data},
  pages     = {1195--1206},
  year      = {2008}
}

@inproceedings{abadi2007materialization,
  author    = {Abadi, Daniel J. and Myers, Daniel S. and DeWitt, David J. and Madden, Samuel R.},
  title     = {Materialization Strategies in a Column-Oriented {DBMS}},
  booktitle = {Proceedings of the 23rd International Conference on Data Engineering (ICDE)},
  pages     = {466--475},
  year      = {2007}
}

@inproceedings{ailamaki2001pax,
  author    = {Ailamaki, Anastassia and DeWitt, David J. and Hill, Mark D. and Skounakis, Marios},
  title     = {Weaving Relations for Cache Performance},
  booktitle = {Proceedings of the 27th International Conference on Very Large Data Bases (VLDB)},
  pages     = {169--180},
  year      = {2001}
}

@article{denning1968workingset,
  author  = {Denning, Peter J.},
  title   = {The Working Set Model for Program Behavior},
  journal = {Communications of the ACM},
  volume  = {11},
  number  = {5},
  pages   = {323--333},
  year    = {1968}
}

@article{noghabi2017samza,
  author  = {Noghabi, Shadi A. and Paramasivam, Kartik and Pan, Yi and Ramesh, Navina and Bringhurst, Jon and Gupta, Indranil and Campbell, Roy H.},
  title   = {Samza: Stateful Scalable Stream Processing at {LinkedIn}},
  journal = {Proceedings of the VLDB Endowment},
  volume  = {10},
  number  = {12},
  pages   = {1634--1645},
  year    = {2017}
}

@misc{qdrantmultitenancy,
  author = {{Qdrant}},
  howpublished = {\url{https://qdrant.tech/documentation/manage-data/multitenancy/}},
  key = {Qdrant},
  note = {Tenant index (\texttt{is\_tenant}, v1.11) and tiered multitenancy (v1.16). Accessed November 2026},
  title = {Qdrant Documentation: Configure Multitenancy},
  year = {2026}
}

@misc{milvusmultitenancy,
  author = {{Milvus}},
  howpublished = {\url{https://milvus.io/docs/multi_tenancy.md}},
  key = {Milvus},
  note = {Accessed November 2026},
  title = {Milvus Documentation: Implement Multi-tenancy},
  year = {2026}
}

@inproceedings{maharana2024locomo,
  address = {Bangkok, Thailand},
  author = {Maharana, Adyasha and Lee, Dong-Ho and Tulyakov, Sergey and Bansal, Mohit and Barbieri, Francesco and Fang, Yuwei},
  booktitle = {Proceedings of the 62nd Annual Meeting of the Association for Computational Linguistics (Volume 1: Long Papers)},
  doi = {10.18653/v1/2024.acl-long.747},
  pages = {13851--13870},
  publisher = {Association for Computational Linguistics},
  title = {Evaluating Very Long-Term Conversational Memory of {LLM} Agents},
  url = {https://aclanthology.org/2024.acl-long.747/},
  year = {2024}
}

@misc{openviking,
  author = {{Volcengine Viking Team}},
  howpublished = {Software and documentation},
  note = {Accessed November 23, 2026},
  title = {{OpenViking}: The Context Database for {AI} Agents},
  url = {https://github.com/volcengine/OpenViking},
  year = {2026}
}

@inproceedings{wu2025longmemeval,
  author = {Wu, Di and Wang, Hongwei and Yu, Wenhao and Zhang, Yuwei and Chang, Kai-Wei and Yu, Dong},
  booktitle = {Proceedings of the Thirteenth International Conference on Learning Representations},
  title = {{LongMemEval}: Benchmarking Chat Assistants on Long-Term Interactive Memory},
  url = {https://proceedings.iclr.cc/paper_files/paper/2025/hash/d813d324dbf0598bbdc9c8e79740ed01-Abstract-Conference.html},
  year = {2025}
}

@misc{yang2025qwen3,
  archiveprefix = {arXiv},
  author = {Yang, An and Li, Anfeng and Yang, Baosong and Zhang, Beichen and Hui, Binyuan and Zheng, Bo and Yu, Bowen and Gao, Chang and Huang, Chengen and Lv, Chenxu and Zheng, Chujie and Liu, Dayiheng and Zhou, Fan and Huang, Fei and Hu, Feng and Ge, Hao and Wei, Haoran and Lin, Huan and Tang, Jialong and Yang, Jian and Tu, Jianhong and Zhang, Jianwei and Yang, Jianxin and Yang, Jiaxi and Zhou, Jing and Zhou, Jingren and Lin, Junyang and Dang, Kai and Bao, Keqin and Yang, Kexin and Yu, Le and Deng, Lianghao and Li, Mei and Xue, Mingfeng and Li, Mingze and Zhang, Pei and Wang, Peng and Zhu, Qin and Men, Rui and Gao, Ruize and Liu, Shixuan and Luo, Shuang and Li, Tianhao and Tang, Tianyi and Yin, Wenbiao and Ren, Xingzhang and Wang, Xinyu and Zhang, Xinyu and Ren, Xuancheng and Fan, Yang and Su, Yang and Zhang, Yichang and Zhang, Yinger and Wan, Yu and Liu, Yuqiong and Wang, Zekun and Cui, Zeyu and Zhang, Zhenru and Zhou, Zhipeng and Qiu, Zihan},
  doi = {10.48550/arXiv.2505.09388},
  eprint = {2505.09388},
  primaryclass = {cs.CL},
  publisher = {arXiv},
  title = {{Qwen3} Technical Report},
  url = {https://arxiv.org/abs/2505.09388},
  year = {2025}
}

@misc{zhang2025compassjudger2,
  archiveprefix = {arXiv},
  author = {Zhang, Taolin and Cao, Maosong and Lam, Alexander and Zhang, Songyang and Chen, Kai},
  doi = {10.48550/arXiv.2507.09104},
  eprint = {2507.09104},
  primaryclass = {cs.CL},
  publisher = {arXiv},
  title = {{CompassJudger-2}: Towards Generalist Judge Model via Verifiable Rewards},
  url = {https://arxiv.org/abs/2507.09104},
  year = {2025}
}

\end{document}